\documentstyle[11pt,newpasp,twoside,epsf]{article}
\begin{document}
\title{Visualization of Large Multi-Dimensional Datasets}
 \author{Joel Welling}
\affil{Department of Statistics, Carnegie
          Mellon University, Pittsburgh, PA 15213, U.S.A. and Pittsburgh
          Supercomputing Center, Pittsburgh, PA 15213, U.S.A.}
\author{Mark Derthick}
\affil{Human Computer Interaction Institute, Carnegie
          Mellon University, Pittsburgh, PA 15213, U.S.A.}

\index{Visualization}
\index{Interactivity}
\index{Condensed Representations}

\begin{abstract}
Visualization techniques are well developed for many problem domains,
but these systems break down for datasets which are very large or
multidimensional.  Techniques for data which is discrete rather than 
continuous are also less well studied.  Astronomy datasets like the
Sloan Digital Sky Survey are very much in this category.  We propose
the extension of information visualization techniques to these very
large record-oriented datasets.  Specifically, we describe the
possible adaptation of the {\sl Visage} information visualization tool to
terabyte astronomy datasets.
\end{abstract}

\section{Introduction}

How can huge data sources (gigabytes up to terabytes) be quickly and
easily analyzed?  There is no {\sl off-the-shelf} technology for this.
There are devastating computational and statistical difficulties;
manual analysis of such data sources is now passing from being simply
tedious into a new, fundamentally impossible realm where the data
sources are just too large to be assimilated by humans.  The only
alternative is to provide extensive computer support for the process
of discovery.

The focus of this workshop is the challenge posed by the next
generation of large astronomical sky surveys.  Specifically, we
concentrate on the Sloan Digital Sky Survey (SDSS), which will create
over one terabyte of reduced data over the next 5 years: How does one
navigate such a huge, multi-dimensional, dataset?  The techniques of
information visualization and visual discovery may be extended to
such datasets, allowing the scientist to interactively explore and
understand her results in real time.

This work is the product of a collaboration at Carnegie Mellon
University and the University of Pittsburgh, involving astronomers,
experts in traditional scientific visualization and interactive
information visualization, and computer and computational scientists.
Condensed data representations from the data mining and machine
learning community (Nichol et al. in these proceedings) make it
possible to explore these huge datasets interactively.

\section{Visualization Challenges of the Virtual Observatory}

Over the next ten years, we will witness a revolution in how
astrophysical research is performed.  This is primarily due to the
large number of new sky surveys presently underway (or completed) that
are designed to map the Universe to higher sensitivity and resolution
than ever previously envisaged.  We are
quickly approaching the prospect of a Virtual Observatory, where one
can digitally reconstruct the whole sky.  These surveys, and the
virtual observatory, present scientists with a ``gold mine'' that the
next generation of astrophysicists will spend their whole careers
exploring.  

Two cornerstones of the Virtual Observatory are the 2 Micron All Sky
Survey (2MASS) and the optical Sloan Digital Sky Survey (SDSS).  The
2MASS survey (which is 91
released) is a near infrared imaging survey covering the full sky in
three passbands (from 1- 2.2 microns). The SDSS is an imaging and
spectroscopic survey that will cover one quarter of the sky at five
different wavelengths.  Together these two surveys will detect over
200 million objects (galaxies and stars) and from these detections
positions, fluxes, shapes, textures and bitmaps will be extracted. In
addition to the scientific information will be bookkeeping information
that describes the observations themselves, e.g. whether the sky was
cloudy or there were problems with the instrumentation.  We must also
understand these possible systematic uncertainties present within the
data.  The total 2MASS and SDSS surveys are expected to acquire 500 GB
of cataloged attributes and 1 TB of postage stamp images (cutout
images around each detected object) over the next 5 years of
operation.

For each object detected hundreds of attributes will be recorded.  The
size and large dimensionality of these new data sets means that simple
visualization and analysis techniques cannot be applied directly,
because they do not scale effectively.  The questions then become: How
do we quickly determine the important dimensions within such a data
set?  Which dimensions tell us about how galaxies or stars form and
how matter is distributed throughout the individual galaxies or, the
Universe as a whole?  New techniques developed for 2MASS and SDSS
will be applicable to the analyses of all large observational data
sets (such as the all-sky surveys of GALEX, ROSAT, MAP and PLANCK) and
for the visualization of other large physical and biological
experiments.

\section{The Breakdown of Visual Discovery}

Traditional astronomy has relied on the study of small numbers of rare
objects.  With hundreds of millions of objects in the Virtual
Observatory, ``rare'' objects will themselves number in the millions.
Statistical methods become necessary to group objects into classes for 
study.

This situation is opposite to that for which traditional visualization 
methods are best suited.  Large dataset size alone kills
interactivity.  The record-oriented nature of the dataset makes it
difficult to assimilate, because the brain is evolved to understand
continuous systems that exist in three dimensions plus time.
If too broad a view of the data is taken, important details can be
literally too small to see.  Too narrow a view is also disastrous.
It becomes very easy to be distracted by structures that {\sl look}
important.

Supercomputer simulations routinely produce very large datasets.
Standard methods of visualizing such data typically include the
generation of one or more animations (assuming the system evolves in
time), or interactive visualization of tiny subsets of the data.  In
most problem domains the researcher's physical intuition applies, and
can help her to assimilate the evolution of the system.

Large computing facilities are developing interactive visualization
tools for these terabyte-sized datasets.  Direct application of
traditional visualization typically requires a supercomputer, and even 
with one available interactivity is very difficult to attain.  Current 
rendering hardware can draw at most about 10 million polygons per
second (though this number is rising), so interactive drawing of
hundreds of millions of objects is simply impossible.

\section{Desirable Capabilities}

Here are illustrative 
questions that astrophysicists will want to ask of the data:

\begin{itemize}

\item 
A range of counting queries such as:

\begin{itemize}
\item
How many elliptical galaxies, with a redshift above 0.3, are there?
\item
What is the mean and variance of ellipticity among radio galaxies
within clusters of galaxies compared to outside clusters of
galaxies?
\item
How does the distribution of galaxy colors observed
on 1 May 2001 compare to those seen on 1 June 2002?
\end{itemize}

\item 
Sophisticated statistical queries that require the clustering,
classification, regression, filtering and newer probabilistic
inference techniques.

\item 
Visualization requests to answer questions such as: 
\begin{itemize}
\item
Give me a smoothed map of all X-ray detected galaxies.
\item
Display all emission-line galaxies in detected clusters of galaxies;
are they in the center or on the outskirts?
\end{itemize}

\end{itemize}

\section{XGobi, a Simple Tool for Visualizing Multidimensional Data}
\label{xgobisec}
\index{XGobi}

\begin{figure}
\plotone{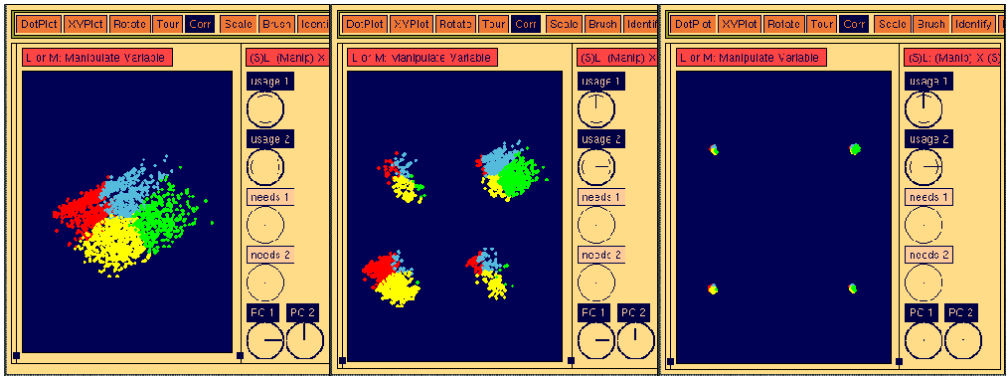}
\caption{The information visualization tool XGobi}
\label{figxgobi}
\end{figure}

XGobi, one of the most widely used information visualization tools,
was written at Lucent Technologies 
\footnote{XGobi is freely available from
http://www.research.att.com/areas/stat/xgobi/.} (Swayne, Cook, \& Buja
1998).  This tool interactively produces arbitrary 3D projections of
n-dimensional data.  For example, figure \ref{figxgobi} shows the
correspondence between data groups which differ in one projection but
are cluster together in another.  XGobi deals only with scalar data,
and is designed for relatively small datasets.

XGobi supports a method called brushing, whereby
the user selects data items in one window and the selection is
propagated to the corresponding items in all other windows.  This has
proven to be a useful technique for mentally integrating multiple views.

\section{Visage, an Information Exploration Tool}
\index{Visage}

At Carnegie Mellon University, in collaboration with Maya Design
Group, we have already developed an interactive data exploration
system called Visage (Kolojejchick, Roth, \& Lucas 1997).  It is
effective for analyzing high dimensional data, but it uses only 2D
visualizations and is limited to small discrete datasets.  Below, we
describe Visage's features, while in the next subsection, we discuss
our plans to generalize these features and expand the capacity of
Visage to handle massive datasets.  We call this new system
TeraVisage.

Visage presents data as graphemes (visual elements) organized by
frames (lightweight nestable windows that impose some visualization
discipline on the graphemes presented within them).
Within a frame, the data are presented by graphemes such as bars, text
labels, marks and gauges.  The visual properties of the graphemes
(e.g., the color or size of a plot point) encode attributes of the
objects they represent (e.g., a galaxy).  While each grapheme stands
for one object, an object may be represented by many graphemes of
varying appearance in different frames.

Some of the basic operations provided by Visage are:

\begin{itemize}

\item 
Drag-and-drop objects from one frame to another, 

\item 
Navigate (drill-down) from an object along a relation creates
graphemes for the related objects,

\item 
Aggregate (roll-up) a set of selected objects to create a new object
with properties computed from its members,

\item 
Brush an object or a set of objects in a choice of colors, as
described in Section \ref{xgobisec},

\item 
Dynamically query by an attribute to render invisible all objects within a
frame whose values for that attribute fall outside a range selected by
a slider widget.

\end{itemize}

The results of these operations depend only on the underlying data
object, and are uniformly applicable in any type of Visage frame.

\subsection{Going to Large Datasets}
\index{TeraVisage}

In this subsection, we discuss the methods by which we propose to
adapt Visage to very large datasets.  Interactive visualization
requires interactive speeds; the challenge is to maintain these speeds
during interrogation of a large (500GB) database.  Our plan is to
maintain a hierarchy of representations and subsets, with a hierarchy
of access speeds.

\begin{figure}
\plotone{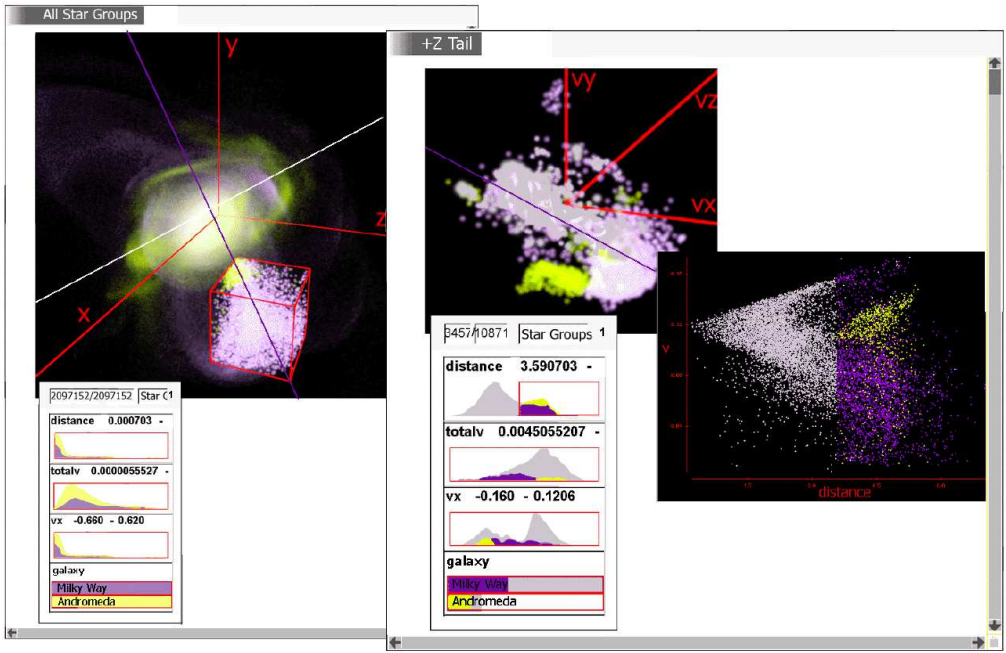}
\caption{Hypothetical TeraVisage operation}
\label{figteravisage}
\end{figure}

Figure \ref{figteravisage} presents our vision of how this would play
out for exploring an astronomical dataset.  The data comes from a
simulation of the coming merger of our Milky Way galaxy with the
Andromeda galaxy\footnote{Welling carried out the simulation using
conditions specified by John Dubinski and code by Lars Hernquist and
others [Mihos, 1998 \#176]; further information is available from
http://www.cita.utoronto.ca/~dubinski/index.html.  An animation of the
merger can be found at http://www.psc.edu/~welling/big-merger.mpg.}.

The dataset we examine here corresponds to the last frame of the
simulation.  In the course of the collision large ``tidal tails'' of
stars, dust, and gas have been drawn from both galaxies.  At the
moment in question the bulk of the galactic matter has formed a
roughly elliptical collision remnant, but two tidal tails remain.  The
matter in one tail is to be selected and examined.

In the left frame labeled ``All Star Groups'', the astronomer has
displayed 7 attributes of all the star groups in the two galaxies.
The 3D visualization shows spatial position (x, y, and z).  The
histograms additionally show the distribution of distance from the
center of the galaxy, magnitude of velocity, x component of velocity,
and original galaxy each star group belonged to.  On the galaxy
histogram the astronomer has brushed the Milky Way star groups purple
and the Andromeda ones yellow.  The relative distributions can be seen
in the other histograms and the density plot.  It is apparent that
star groups that originated in the Milky Way dominate the streamers.
The astronomer selects a group of stars from one of the streamers
using a bounding box.  These stars are brightened in the 3D density
plot.  The brushing and selection operations should take about one
second.

The astronomer then creates a new frame with a density plot showing
the x, y, and z components of the velocities and drags (copies) the
selected star groups there.  She labels the new frame ``+Z Tail''
(right), and copies the histograms as well.  Appropriate condensed
representations should allow this operation to be carried out in about
10 seconds.  Being interested in the relationship between the distance
from the center and the magnitude of the velocity, she creates a 2D
scatter plot of those variables.  By moving the distance slider, she
controls which stars remain fully visible and which are grayed out.
Feedback from slider changes should occur in roughly 100ms.

It is easy to see how to extend most of the graphemes of Visage to
large datasets.  Sets of individual marks become density distributions
in some space, and are selected by selecting regions in the space
and/or using DQ filters.  Gauges remain gauges, but now represent
statistical summaries of subsets of the data.  Text labels remain, but
become labels for sub-aggregates found in traversing kd-trees or in
more sophisticated models.  Many confusing issues remain, however.  

\section{ Interactive Visualization Through Condensed Representations }

In addition to facilitating machine learning and database operations,
appropriate condensed representations can accelerate such tasks as
volume rendering by allowing data items to be grouped and drawn
together if they are similar to within specified error bounds.  A
large dataset can easily have more data elements than there are pixels
on the display, and there is no sense rendering at a level of detail
which will be invisible to the user.  These hierarchical rendering
methods are well studied in computer graphics and map well to
condensed representations for knowledge discovery.  For example,
see be Laur \& Hanrahan (1991).


\begin{references}

\reference Kolojejchick, J. A., Roth, S. F., \& Lucas, P. 1997, in
Computer Graphics and Applications, Vol. 17 No 4,
{\sl Information Appliances and Tools in Visage}, 32,
http://www.cs.cmu.edu/~sage/PDF/Appliances.pdf

\reference  Laur \& Hanrahan 1991 in Computer Graphics, Vol. 25 No. 4,
{\sl Hierarchical Splatting: A Progressive Refinement Algorithm for 
Volume Rendering}, 285

\reference Swayne, D. F., Cook, D., \& Buja, A. 1998,
	in Journal of Computational and Graphical Statistics 7 (1),
	{\sl XGobi: Interactive Dynamic Data Visualization in the X
	Window System}

\end{references}
\end{document}